\documentclass[journal=jpcbfk,manuscript=article,layout=twocolumn]{achemso}

\usepackage[utf8]{inputenc}
\usepackage[T1]{fontenc}
\usepackage[final]{microtype}

\usepackage[american]{babel}
\usepackage{graphicx}
\usepackage{booktabs}

\usepackage[version=4]{mhchem}
\usepackage{siunitx}
\DeclareSIUnit\angstrom{\text {Å}}
\author{Rinto Thomas}
\affiliation[UMR]{Fachbereich Chemie, Philipps-Universität Marburg, 35032 Marburg, Germany}
\altaffiliation{R.T. and P.R.P. contributed equally to this work.}
\author{Praveen Ranganath Prabhakar}
\affiliation[UCI]{Department of Chemistry, University of California, Irvine, Irvine, California, 92697 United States}
\altaffiliation{R.T. and P.R.P. contributed equally to this work.}
\author{Douglas J. Tobias}
\affiliation[UCI]{Department of Chemistry, University of California, Irvine, Irvine, California, 92697 United States}
\email{dtobias@uci.edu}
\author{Michael von Domaros}
\affiliation[UMR]{Fachbereich Chemie, Philipps-Universität Marburg, 35032 Marburg, Germany}
\email{mvondomaros@uni-marburg.de}

\title
  {Insights into Dermal Permeation of Skin Oil Oxidation Products from Enhanced Sampling Molecular Dynamics Simulation}

\begin{document}

\begin{tocentry}
    \includegraphics[width=3.25in]{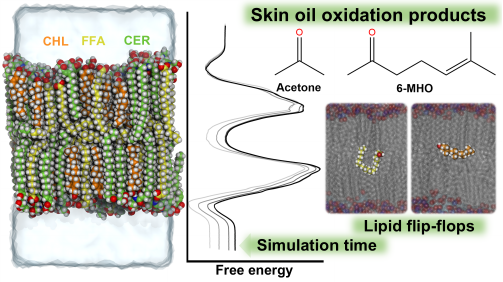}
\end{tocentry}

\begin{abstract}
    The oxidation of human sebum, a lipid mixture covering our skin, generates a range of volatile and semi-volatile carbonyl compounds that contribute largely to indoor air pollution in crowded environments. 
    Kinetic models have been developed to gain a deeper understanding of this complex multiphase chemistry, but they rely partially on rough estimates of kinetic and thermodynamic parameters, especially those describing skin permeation. Here, we employ atomistic molecular dynamics simulations to study the translocation of selected skin oil oxidation products through a model stratum corneum membrane. We find these simulations to be non-trivial, requiring extensive sampling with up to microsecond simulation times, in spite of employing enhanced sampling techniques. We identify the high degree of order and stochastic, long-lived temporal asymmetries in the membrane structure as the leading causes for the slow convergence of the free energy computations.
    We demonstrate that statistical errors due to insufficient sampling are substantial and propagate to membrane permeabilities.
    These errors are independent of the enhanced sampling technique employed and very likely independent of the precise membrane model.
\end{abstract}

\section{Introduction}

Most people live, sleep, work, or learn indoors, and travel or commute in closed vehicles.
Thus, most humans spend the vast majority of their time, usually estimated to be $\sim \SI{90}{\percent}$, indoors.
Therefore, it is natural to question the health aspects of this practice and to study how human presence and activities affect indoor air quality.
This is the main motivation for the field of indoor chemistry, which has evolved over the past decades from a novel branch of atmospheric chemistry to its own field, attracting scientists from a multitude of disciplines.\cite{Weschler.Weschler.2011.ChemistryIndoorEnvironments,Weschler.Carslaw.2018.IndoorChemistry,Beko.Querol.2020.PresentFutureIndoor,Abbatt.Wang.2020.AtmosphericChemistryIndoor,Carslaw.Shaw.2024.NewFrameworkIndoor}

Reaction conditions in indoor environments, such as temperature or relative humidity, are usually markedly different than outdoors, but the most striking differences are the large surface-to-volume ratio and the chemical variety of both surfaces and compounds found indoors.\cite{Ault.Xiong.2020.IndoorSurfaceChemistry,VonDomaros.Tobias.2024.IndoorSurfacesReview}
Previous research has shown that not only human activities, such as cooking, cleaning, or personal care, transform the composition of indoor air,\cite{K.Farmer.E.Vance.2019.IndoorAirSources} but also the mere presence of people in indoor environments.\cite{Kruza.Carslaw.2019.HowBreathSkin}

\begin{figure*}[htbp]
    \centering
    \includegraphics[width=\textwidth]{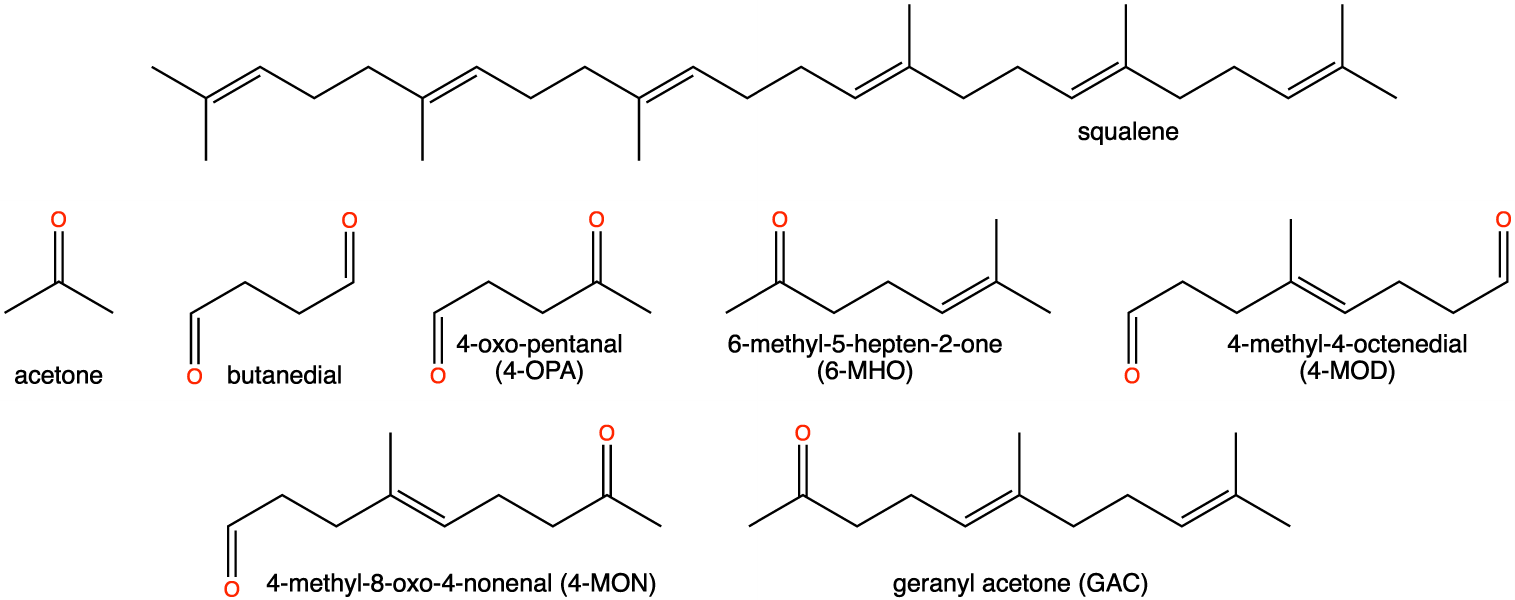}
    \caption{Squalene and its volatile and semivolatile ozonolysis products.}
    \label{fig:structures}
\end{figure*}

A prominent example is the oxidation of the oily-to-waxy mixture of lipids covering our skin, the sebum, through atmospheric oxidants, notably ozone.
This chemistry has been shown to contribute greatly to indoor air pollution in crowded indoor environments, such as large offices, schools, or airplanes.\cite{Weschler.Nazaroff.2007,Wisthaler.Weschler.2010,Weschler.Nazaroff.2023.HumanSkinOil}
Figure~\ref{fig:structures} shows ozonolysis reaction products of squalene, a highly unsaturated skin lipid, which may be oxidized at any of its six double bonds, giving rise to a range of volatile and semivolatile mono- and dicarbonyls, some of which may evoke irritant and allergic responses, explaining some of the adverse health effects of poor indoor air quality.\cite{Anderson.Meade.2012.IrritancyAllergicResponses,Wolkoff.Nielsen.2013.HumanReferenceValues,Beko.Spengler.2015.ImpactCabinOzone,Wolkoff.Wolkoff.2018.IndoorAirHumidity}

Given the high complexity of skin oil oxidation and its impact on indoor air, as well as the ethical considerations of experimenting with human participants, models have emerged to gain a better understanding of this chemistry.
Of particular note is the kinetic multilayer model of the surface and bulk chemistry of skin (KM-SUB-Skin),\cite{Lakey.Shiraiwa.2017.ChemicalKineticsMultiphase} which resolves the mass transport
and chemical reactions of ozone and skin oil oxidation products through various layers of the skin, the gas phase, and the corresponding interfaces, giving access to spatial concentration profiles on millimeter to micrometer length scales and temporal concentration profiles on time scales up to hours.
A recent extension of the model, KM-SUB-Skin-Clothing, includes additional layers of clothing and the corresponding boundary layers.\cite{Lakey.Shiraiwa.2019.ImpactClothingOzone}
The results from such models may be fed into computational fluid dynamics simulations for a room-scale perspective.\cite{Lakey.Shiraiwa.2019.ImpactClothingOzone}

Evaporation into indoor air is only one of two major pathways out of the skin for sebum oxidation products, the other being dermal absorption and transport into blood vessels.
The respective fluxes depend critically on modeling choices, notably on parameters describing the partitioning and the transport of species throughout the skin.\cite{Lakey.Shiraiwa.2017.ChemicalKineticsMultiphase}
Unfortunately, most of these parameters are unavailable experimentally and need to be computed or estimated. In the original formulation of KM-SUB-Skin, parameters were estimated using empirical models parameterized against databases of drugs and drug-like compounds,\cite{Wang.Nitsche.2007.MultiphaseMicroscopicDiffusion} and dependent on molecular data, such as molar mass, octanol--water partition coefficient, and volume. The chosen approach completely neglects the chemical diversity of both the solutes and the skin lipids and the microscopic structure of the skin.
It is also unclear whether the parameterization employed by these models is valid for the skin oil oxidation products shown in Figure~\ref{fig:structures}.

We have previously used molecular dynamics~(MD) simulations to study the skin-oil--gas interface in order to estimate parameters required by kinetic models, and to gain an atomistically resolved mechanistic understanding of this complex chemistry.\cite{VonDomaros.Tobias.2020.MultiscaleModelingHuman} 
Combined with kinetic modeling and computational fluid dynamics, a comprehensive, multiscale physical modeling approach emerges.\cite{Shiraiwa.Won.2019.ModellingConsortiumChemistry}

In this article, we extend our MD simulations to the prediction of skin permeabilities, highlighting the challenges encountered in the process. 
We note that other computational approaches exist, typically quantitative structure–activity relationships (QSAR) and related models, but their discussion would go beyond the scope of this work.\cite{Pecoraro.Traynor.2019}

The challenges in predicting skin permeabilities from MD simulations start with the choice of a model system.
In principle, human skin is a complex, multilayered organ of millimeter thickness, that cannot be described in full detail by atomistic techniques.
Thus, several approximations need to be made in the hope of finding a representative model system that is accessible on MD simulation length and time scales.

The barrier function of the skin lies almost completely in its topmost layer, the stratum corneum (SC), historically described as a brick-and-mortar-like arrangement of several layers of cells (corneocytes) embedded in a lipid matrix.\cite{Matsui.Amagai.2015}
Of all SC components, the lipid matrix is the only one that extends fully through the stratum corneum and therefore is of great interest in the study of SC permeation processes.
It is a roughly equimolar mixture of cholesterol, ceramides, and free fatty acids, with a high degree of chemical variability among the different classes of lipids, and variations in composition with body parts, age, sex, and other individual factors. \cite{Weerheim.Ponec.2001.DeterminationStratumCorneum}
X-ray diffraction measurements have revealed the presence of two coexisting lamellar phases with repeat units of \num{6} and \SI{13}{nm}, termed short (SPP) and long periodicity phase (LPP), respectively.\cite{Bouwstra.Bras.1991.StructuralInvestigationsHuman}
The atomic structure of the LPP is still largely unknown, and a number of completely different model systems have been developed over the years.\cite{Schmitt.Neubert.2020.StateArtStratum,Bouwstra.Gooris.2023.SkinBarrierExtraordinary,Shamaprasad.McCabe.2024.PhaseBehaviorSkinbarrier}
The repeat units of the SPP, on the other hand, are consistent with a simple bilayer structure, though the conformation of the double-chained ceramides (fully extended vs. hairpin) is still a matter of debate; most likely, there is a dynamical equilibrium between multiple conformations.\cite{Schmitt.Neubert.2020.StateArtStratum}

For the sake of simplicity, most SC simulation studies focus on fully hydrated bilayer membranes, in which ceramides cannot adopt fully extended conformations, though more recent studies attempt to rectify this approximation.\cite{Shamaprasad.McCabe.2022.MultiscaleSimulationTernary,Shamaprasad.MCabe.2022.UsingMolecularSimulation}
Once membrane permeabilities are known, transport models\cite{Mitragotri.Roberts.2011,Frasch.Barbero.2013,Wang.Naegel.2023} can be used to extrapolate them to the full SC.

In this article, we study the permeation of water and two squalene ozonolysis products, acetone and 6-methyl-5-hepten-2-one (6-MHO), through model SC SPP bilayer membranes.
We note that despite being volatile, 6-MHO is one of the few reaction products that permeates significantly through the skin according to the KM-SUB-Skin model.\cite{Lakey.Shiraiwa.2017.ChemicalKineticsMultiphase}
We do not seek to address the modeling challenges described above, noting that the employed model is rather simplistic in comparison to other SC models.
Instead, we wish to highlight methodological challenges in the prediction of free energy profiles describing the thermodynamic aspects of permeation, which become apparent in simple SC models and are very likely present in more complex systems as well.

\section{Computational Details}

\subsection{Model system and force field}

\begin{figure}[htb]
    \centering
    \includegraphics[width=\columnwidth]{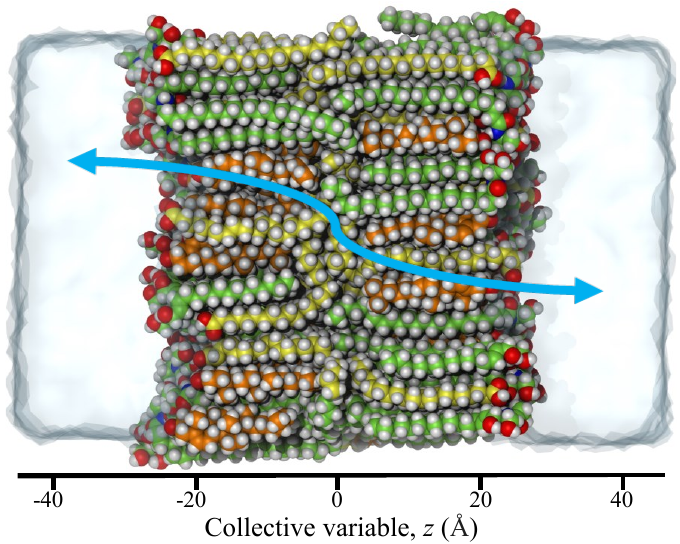}
    \caption{Schematic representation of a permeation path through the SC model employed in this study. Lipid carbon atoms are colored orange (CHL), green (CER) and yellow (FFA).
    Oxygen atoms are shown in red and nitrogen in blue.
    The water phase is represented as a transparent surface. The collective variable (CV) is the projection of the path onto the membrane normal, with $z = 0$ corresponding to the center of the membrane.}
    \label{fig:system}
\end{figure}

We studied the fully hydrated SPP bilayer model developed by Wang and Klauda.\cite{wang_models_2018}
This equimolar, symmetric, three-component model comprises 68~N-lignoceroylsphingosine ceramides~(CER), 66~cholesterol~(CHL) and 66~protonated lignoceric acid molecules~(FFA), solvated by 4800~water molecules (Figure~\ref{fig:system}).

We frequently compare our results for the multicomponent SC model with a much simpler single-component fluid lipid bilayer composed of 1-palmitoyl-2-oleoyl-sn-glycero-3-phosphocholine (POPC).
The POPC system is composed of 128~POPC molecules solvated by 4040~water molecules. The initial POPC membrane structure was obtained from the CHARMM-GUI website.\cite{jo_charmm-gui_2008}

SC and POPC lipids were modeled using the CHARMM36 lipid force field.\cite{klauda_update_2010}
Acetone is one of the CGenFF force field\cite{vanommeslaeghe_charmm_2010} model compounds and the parameters and charges included with version~4.6 of this force field were used.
For 6-MHO, values were automatically assigned by the CGenFF program (version~4.0).
A parameter penalty score of \num{3.500} and a charge penalty score of \num{6.377} were obtained. In both cases, values less than 10 indicate a good match with similar compounds in the CGenFF force field.
The CHARMM-modified TIP3P water model was used in all simulations.\cite{jorgensen_comparison_1983,MacKerell.Karplus.1998.AllAtomEmpiricalPotential}

To study permeation, we calculated the free energy profiles for translocating molecules across the membrane.
For this purpose, we used a collective variable (CV), $z$, measuring the distance of a solute toward the membrane normal~(Figure~\ref{fig:system}).

\subsection{Simulation protocol}

\subsubsection{Common simulation parameters}

The equations of motion were integrated using a multiple timestep algorithm (r-RESPA),\cite{humphreys_multiple-time-step_1994} with a \SI{4}{\fs} time step for electrostatic and a \SI{2}{\fs} time step for all other interactions.
The water molecules were kept rigid by the non-iterative SETTLE algorithm,\cite{miyamoto_settle_1992} and all other bonds to hydrogen atoms were constrained by the SHAKE-H algorithm.\cite{ryckaert_numerical_1977}
All simulations were performed at the physiological skin temperature of $T =\qty{305.15}{\kelvin}$, which was controlled by a stochastic velocity scaling thermostat \cite{bussi_canonical_2007} with a time constant of \SI{1}{\ps}.
Short-range non-bonded forces were smoothly shifted to zero between \num{10} and \qty{12}{\angstrom} and the smooth particle mesh Ewald (SPME)\cite{essmann_smooth_1995} method with a \qty{1}{\angstrom} grid and with sixth-order spline interpolation was employed to compute long-range electrostatic interactions.

Unless otherwise noted, all simulations were performed using NAMD~2.14.\cite{phillips_scalable_2005,phillips_scalable_2020}
Molecular representations were rendered using VMD~1.9.4\cite{humphrey_vmd_1996} and analyses were performed with VMD~1.9.4 and MDAnalysis~2.7.0.\cite{michaud-agrawal_mdanalysis_2011, gowers_mdanalysis_2016}

\subsubsection{Unbiased simulations}

The solutes were inserted into the center of the aqueous region by removing water molecules within a cutoff distance of \qty{2}{\angstrom} and the energy of the system was minimized by performing \num{50000} steepest descent steps.
The system was then equilibrated for \qty{10}{ns} at a constant pressure of $P = \SI{1}{atm}$ by applying a Nosé–Hoover Langevin piston with a piston oscillation time of \SI{100}{\fs} and a damping time scale of \SI{50}{\fs}.\cite{martyna_constant_1994,feller_constant_1995}
The pressure control was anisotropic, with independent pressure control in the direction of the membrane normal.
The cell size of the system was then set to the average size observed during the constant-pressure run, and the system was equilibrated for another \qty{50}{ns} at constant volume. 
Unbiased properties of the membranes, such as lipid order parameters, were evaluated from subsequent \qty{100}{\nano\second} production runs. 

\subsubsection{Umbrella sampling}

We performed $N_\text{win}=91$ independent simulations with harmonic potentials 
\begin{align}
    V_n(z)=\frac{K}{2}\left(z-z_n\right)^2 && (n=0\dots{}N_\mathsf{win}-1)
\end{align}
restraining the solute to stay within a given window of the collective variable.
The bias potentials were centered at
$z_n = z_0 + n\Delta{}z$, with $z_0=\SI{-45}{\angstrom}$ and $\Delta{}z=\SI{1}{\angstrom}$.
The force constant $K$ was set to \SI{2.5}{kcal\per\mole}.
Each system was allowed to equilibrate for \SI{5}{ns}, the system with water as permeant was simulated for \SI{300}{ns} per window, and the acetone and 6-MHO systems were simulated for \SI{600}{ns} per window.

To generate the initial configurations for the umbrella sampling (US),\cite{torrie_nonphysical_1977} the solute was first pulled from one side of the membrane to the other, by moving the center of the harmonic potential in steps of $\Delta{}z$ every \SI{10}{\pico\second}. Snapshots closest to target centers were identified and used in the simulations described above.

We followed a similar protocol for the POPC systems, but with $N_\text{win}=65$ and $z_0=\SI{-32}{\angstrom}$. Each POPC system was simulated for \SI{150}{ps} per window.

US was performed with NAMD 3.0alpha13, which natively supports this biasing technique. The CV was saved every \SI{100}{fs} for post-simulation analysis.
The weighted histogram analysis method (WHAM)\cite{kumar_weighted_1992,rochesterWHAMx2013} was used to calculate the free energy (FE) profile of the solute.
The error bars shown in the FE profiles represent $\SI{95}{\percent}$ confidence intervals and were estimated as described by Zhu and Hummer.\cite{Zhu.Hummer.2011.ConvergenceErrorEstimation}
The required standard errors of the mean of the collective variable were estimated using an automated blocking method.\cite{Jonsson.Jonsson.2018.StandardErrorEstimation} 
Symmetrized FE profiles and confidence intervals were obtained by treating both halves of the membrane as statically independent samples.
More details on the computation of error bars are given in the SI.

\subsubsection{Well-tempered metadynamics}

We also performed well-tempered metadynamics (WTM) simulations.\cite{barducci_well-tempered_2008}
In the SC simulations, we set the bias factor, $\gamma = \left(T+\Delta{}T\right)/T$, to 20 for water and acetone and to 15 for 6-MHO.
This corresponds to temperature boosts $\Delta{}T$ of approximately \SI{5800}{\kelvin} and \SI{4300}{\kelvin}, respectively.
In POPC simulations, a bias factor of 8 ($\Delta{T} \approx \SI{2100}{\kelvin}$) was used for all solutes.
In all WTM simulations, the hill widths were set to \qty{2}{\angstrom} and hills of the height \SI{0.1}{kcal\per\mol} were deposited at a rate of \SI{1}{\per\ps}.
The trajectory files were saved every \qty{0.1}{\nano\second} for post-simulation analysis.
WTM simulations were performed with NAMD~2.14 and the colvars library.\cite{fiorin_using_2013} 
The standard error of the mean of the FE profiles was estimated by applying an automated blocking method\cite{Jonsson.Jonsson.2018.StandardErrorEstimation} to snapshots of the FE energy profiles saved every \SI{2}{\nano\second}.
Once again, error bars shown correspond to \SI{95}{\percent} confidence intervals.
More details on the computation of error bars are given in the SI.

\section{Results and Discussion}

\subsection{Convergence to symmetric free energy profiles is unexpectedly slow}

Because partitioning of molecules between the solvent and the membrane is a critical step in understanding membrane permeation, we computed Helmholtz free energy (FE) profiles (also known as potentials of mean force, PMFs) describing membrane translocation for all investigated species.

\begin{figure*}[htbp]
     \centering
     \includegraphics[width=\textwidth]{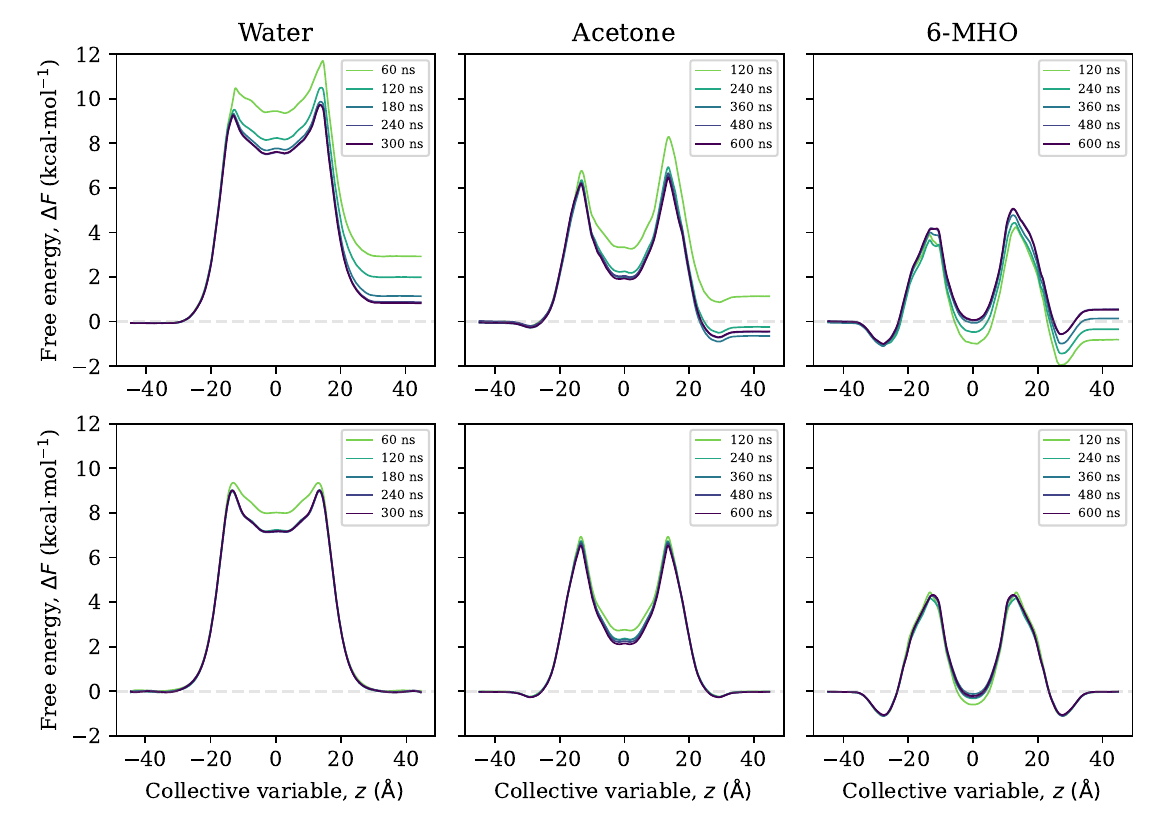}
     \caption{Top row: Convergence of FE profiles toward symmetry for US. Bottom row: Convergence of the corresponding force-symmetrized FE profiles toward their final values. All simulation times are per window. Total simulation times can be obtained by multiplying with the number of windows ($N_\text{win} = 91$). They are also reported in Table~\ref{tab:simulation_times}.}
     \label{fig:us_convergence_fe}
\end{figure*}

\begin{figure*}[htbp]
     \centering
     \includegraphics[width=\textwidth]{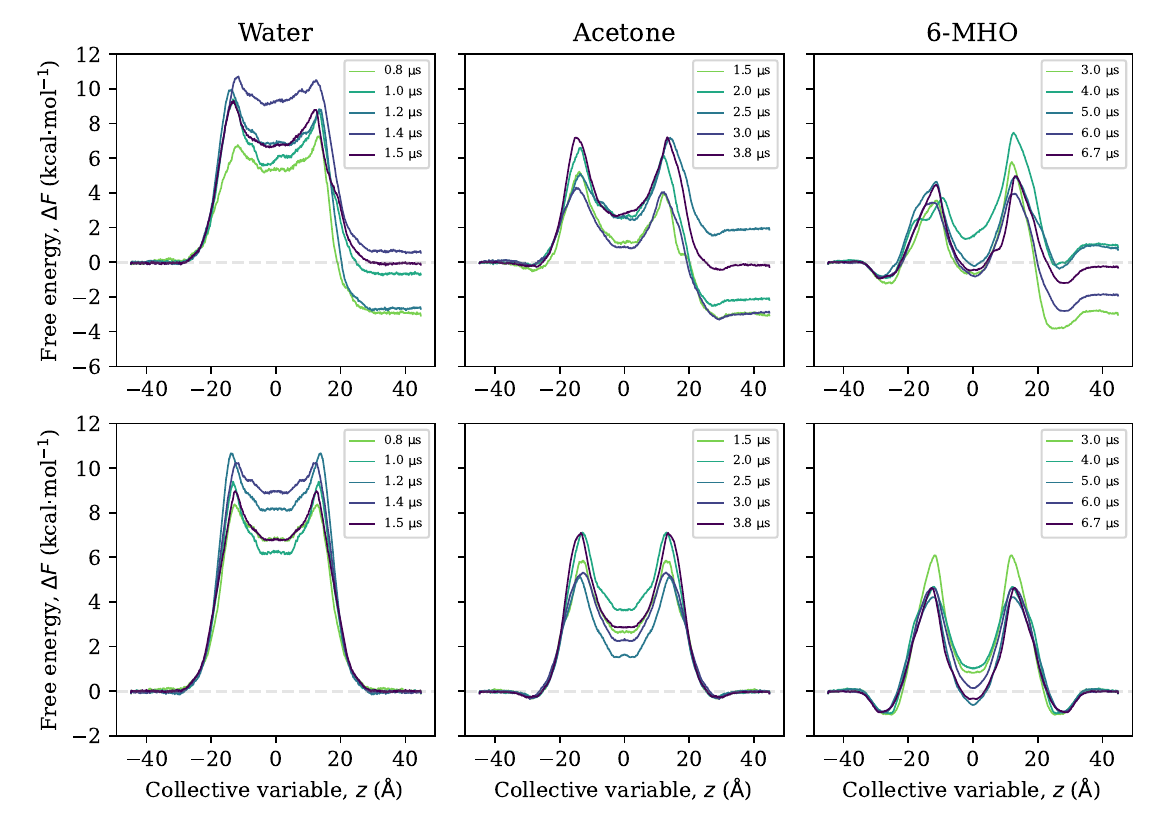}
     \caption{Top row: Convergence of FE profiles toward symmetry for WTM. Bottom row: Convergence of the corresponding force-symmetrized FE profiles toward their final values.}
     \label{fig:wtm_convergence_fe}
\end{figure*}

\begin{table}[tb]
\centering
    \caption{Simulation times in \si{\micro\second} required to reach near-symmetry in FE profiles. For US, we report both simulation times per window and totals.}
  \begin{tabular*}{\columnwidth}{@{\extracolsep{\fill}}lrrr}
    \toprule
    System & US & US & WTM \\ 
           & (window) & (total) \\
    \midrule
    SC/Water     & 300 & 27.3 & 1.5 \\ 
    SC/Acetone   & 600 & 54.6 & 3.8 \\
    SC/6-MHO     & 600 & 54.6 & 6.7 \\
    POPC/Acetone & 150 &  9.8 & 0.8 \\
    POPC/Water   & 150 &  9.8 & 0.7 \\
    \bottomrule
    \label{tab:simulation_times}
  \end{tabular*}
\end{table}

In the absence of external fields, the orientation of the membrane normal is arbitrary, and profiles along this axis must be symmetric for bilayers with identical leaflet compositions.
For simple profiles, such as the density profile (Figure~S2), such a symmetry is easily observed.
Thus, many authors choose to restrict simulations to only one-half of the system for computational simplicity.
However, for FE profiles, this was shown to conceal underlying sampling issues, which are not properly taken into account by common error estimation techniques.\cite{pokhrel_free_2018}
Consequently, we estimated FE profiles for the complete system and decided to run our simulations until we reached close-to-symmetric profiles.

We found this convergence to be exceptionally slow, which we demonstrate in the top row of Figure~\ref{fig:us_convergence_fe} for US---where total simulation times easily reach tens of \si{\micro\second} (Table~\ref{tab:simulation_times}).
In terms of total simulation time, WTM performs better, but still requires a significant simulation effort of the order of \si{\micro\second} (Figure~\ref{fig:wtm_convergence_fe}, top row; Table~\ref{tab:simulation_times}).
We note that the difference between the free energy plateaus in the aqueous phase may easily exceed \SI{1}{kcal\per\mole} if the simulations are insufficiently long.

Realizing that the optimal performance for both enhanced sampling techniques employed depends critically on the proper choice of parameters, such as the placement of harmonic potentials in US or hill widths and heights in WTM, we chose to benchmark and compare our simulation results for the SC against a POPC system, which represents a much simpler, single component fluid membrane that has been extensively studied in the context of membrane permeation.\cite{Awoonor-Williams.Rowley.2016.MolecularSimulationNonfacilitated,Shinoda.Shinoda.2016.PermeabilityLipidMembranes,Frallicciardi.Poolman.2022.MembraneThicknessLipid}
Our results for water and acetone permeants demonstrate that the convergence of the FE profiles toward symmetry is significantly slower in the SC membrane than in POPC (Table~\ref{tab:simulation_times}, Figures~S3,~S4) and that the slow convergence is a feature of the SC system and not a shortcoming of the employed sampling method.

As an alternative to simulating until symmetry has been reached, one could also force-symmetrize the FE profiles and check for convergence with respect to simulation time.
For US, this procedure leads to much faster convergence (Figure~\ref{fig:us_convergence_fe}, bottom row), and we will show later that this translates directly into the convergence of permeabilities with respect to simulation time.
For WTM, on the other hand, the series of force-symmetrized FE profiles has only converged for 6-MHO, but not for water and acetone (Figure~\ref{fig:wtm_convergence_fe}, bottom row).

\begin{figure}[tb]
     \centering
     \includegraphics[width=\columnwidth]{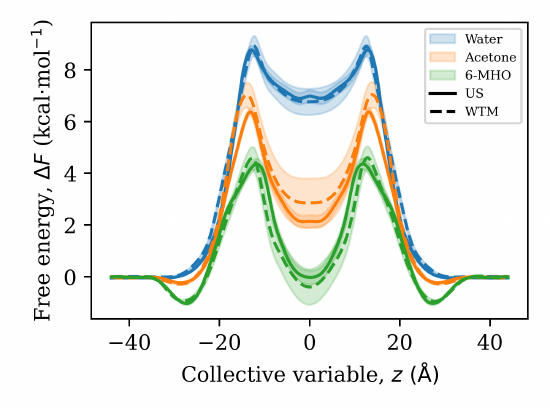}
     \caption{Final, force-symmetrized FE profiles for all investigated solutes in the SC model. Lines and error bars (representing \SI{95}{\percent} confidence intervals) have been slightly smoothed for visual clarity. The raw data is shown in Figure~S5.}
     \label{fig:fe_profiles}
\end{figure}

For a rigorous comparison between US and WTM results, we evaluated the statistical accuracy of our final force-symmetrized free energy profiles, shown in Figure~\ref{fig:fe_profiles} for SC and Figure~S6 for POPC.
The \SI{95}{\percent} confidence intervals for US and WTM overlap for water and 6-MHO, indicating that there are no significant methodological differences in terms of free energy.
We also find a good agreement between our results for water (both in SC and POPC) and previous studies of similar systems.\cite{piasentin_evaluation_2021,sajadi_simulations_2018,kramer_membrane_2020}

\begin{table}[tb]
\centering
    \caption{Barrier heights with respect to the point of reference in the aqueous phase and \SI{95}{\percent} confidence intervals in \si{kcal\per\mol}.}
  \begin{tabular*}{\columnwidth}{@{\extracolsep{\fill}}llrr}
    \toprule
    System & Solute & US & WTM \\ 
    \midrule
    SC      & Water     & $8.75 \pm 0.17$ & $8.96 \pm 0.45$ \\
    SC      & Acetone   & $6.38 \pm 0.15$ & $7.09 \pm 0.54$ \\
    SC      & 6-MHO     & $4.35 \pm 0.20$ & $4.62 \pm 0.48$ \\
    POPC    & Acetone   & $2.60 \pm 0.15$ & $2.59 \pm 0.29$ \\
    POPC    & Water     & $6.65 \pm 0.15$ & $7.04 \pm 0.36$ \\
    \bottomrule
    \label{tab:barrier_heights}
  \end{tabular*}
\end{table}

Acetone FE profiles overlap in most regions of the CV space, but not in the critical barrier region (Table~\ref{tab:barrier_heights}).
These barriers are a simple indicator for permeability, and we will confirm later that the methodological differences described here translate into significant differences in the permeabilities as well.
In the next section, we first investigate the molecular origins of the slow convergence behavior observed for the FE profiles.

\begin{figure*}[htbp]
 \centering
 \includegraphics[width=\textwidth]{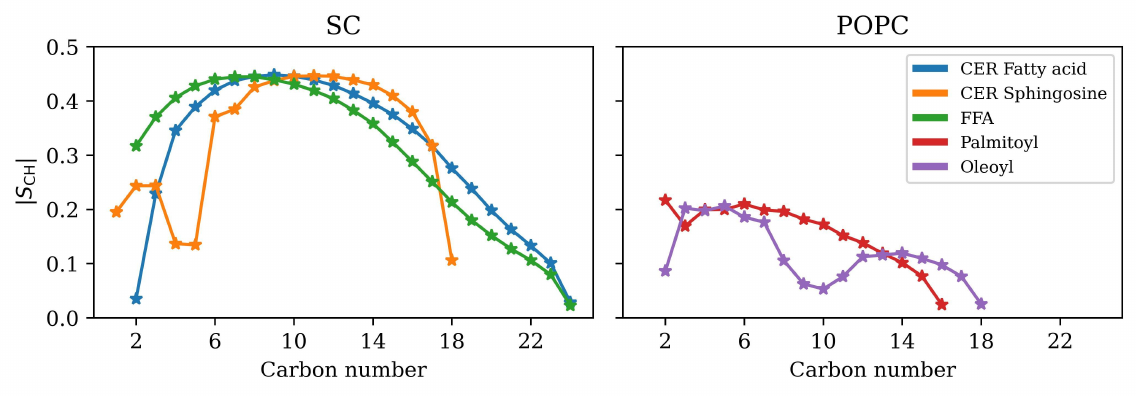}
 \caption{Carbon--hydrogen order parameters for the acyl and free fatty acid chains found in SC and POPC lipids.}
 \label{fig:order}
\end{figure*}

\subsection{SC membranes are highly ordered}

SC membranes are often described as more solid-like than typical fluid membranes, such as those composed of POPC.\cite{das_water_2009,das_physics_2016}
High order and low lipid mobility could certainly be a reason for the slow convergence of the FE profiles demonstrated above.
In this section, we study the structure of the membrane by computing carbon--hydrogen (or carbon--deuterium) order parameters, which are routinely used to characterize the order of acyl chains found in biological membranes and can also be determined experimentally using NMR spectroscopy.\cite{piggot_calculation_2017,seelig_effect_1977}

For a tagged \ce{C-H} bond,
\begin{equation}
    S_{\ce{CH}} = \frac{\left<3 \cos^2 \theta - 1\right>}{2}\,,
\end{equation}
where $\theta$ is the angle between the bond and the membrane normal.
We evaluated this quantity for all unique \ce{C-H} bonds in the acyl chains of the lipids, averaging over equivalent bonds, if possible.

When discussing these order parameters, it is customary to refer to absolute values. Bonds that are perfectly aligned with the membrane normal ($\theta = 0$) exhibit the maximum value, $|S_{\ce{CH}}|=0.5$. The order parameter becomes zero for uniformly distributed bonds or for perfectly ordered bonds oriented at the "magic" angle $\theta = \SI{54.7}{\degree}\ (\cos^2\theta = \frac{1}{3})$.

Our results for the SC system (Figure~\ref{fig:order}, left panel) agree quantitatively with the work of Wang et al.\cite{wang_models_2018}
Those for the POPC system (Figure~\ref{fig:order}, right panel) agree well with the results reported by Piggot et al.\cite{piggot_calculation_2017}
In both systems, the absolute values of the order parameter are greater in the upper and middle regions of the acyl chains compared to those in the lipid tails, which is the result of higher conformational order in these regions.
The presence of double bonds (\ce{C{4}=C{5}} in the sphingosine chains of SC ceramides and \ce{C{9}=C{10}} in the POPC oleyol chain) leads to a dip in $|S_{\ce{CH}}|$ values, as these bonds are aligned around the magic angle $\theta = \SI{54.7}{\degree}$ discussed above (see SI for a more detailed analysis).

When comparing both systems, we see that the internal organization of the SC membrane is markedly different from that of POPC.
The order parameters for carbon atoms in the headgroup region reach absolute values of up to 0.4 in the SC system, whereas the corresponding order parameters for POPC are about \SI{50}{\percent} smaller.
Furthermore, the order parameters for SC lipids are approximately constant over a range of carbon atoms up to \ce{C{13}}, whereas they decay rapidly in POPC lipids.

From this analysis of carbon--hydrogen order parameters, we clearly see that the SC membrane is more ordered and solid-like than the POPC membrane. These structural differences can impact the permeation of solutes and consequentially the free energies corresponding to their partition across different regions of the membrane.   

\subsection{Long-lived asymmetries arise due to lipid flipping in the SC membrane}

\begin{figure}[htb]
 \centering
 \includegraphics[width=\columnwidth]{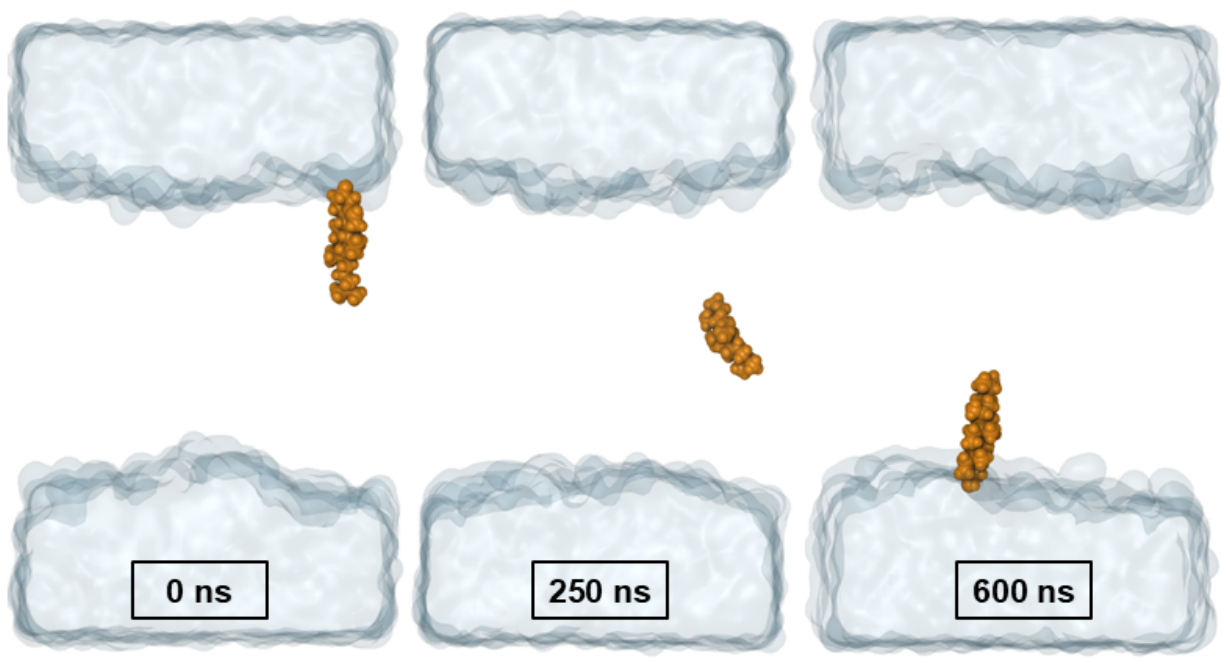}
 \caption{Snapshots of a tagged cholesterol molecule flipping to the opposite layer of the SC membrane in one of the US simulations.}
 \label{fig:flipping_example}
\end{figure}

Visual inspection of the trajectories revealed the occasional flipping of lipids, that is, the departure of the lipids from their average position and orientation along the membrane normal and their motion into the center of the membrane; see Figure~\ref{fig:flipping_example} for an example.
Flipping was observed only for SC lipids and not in POPC,
and occurred in both unbiased and biased (US, WTM) simulations.
Such events lead to rare, but long-lived changes in the membrane structure, which could explain the observed asymmetries in the FE profiles.

\subsubsection{Flipping events can be tracked by following tilt angles}

\begin{figure}[htb]
 \centering
 \includegraphics[width=0.9\columnwidth]{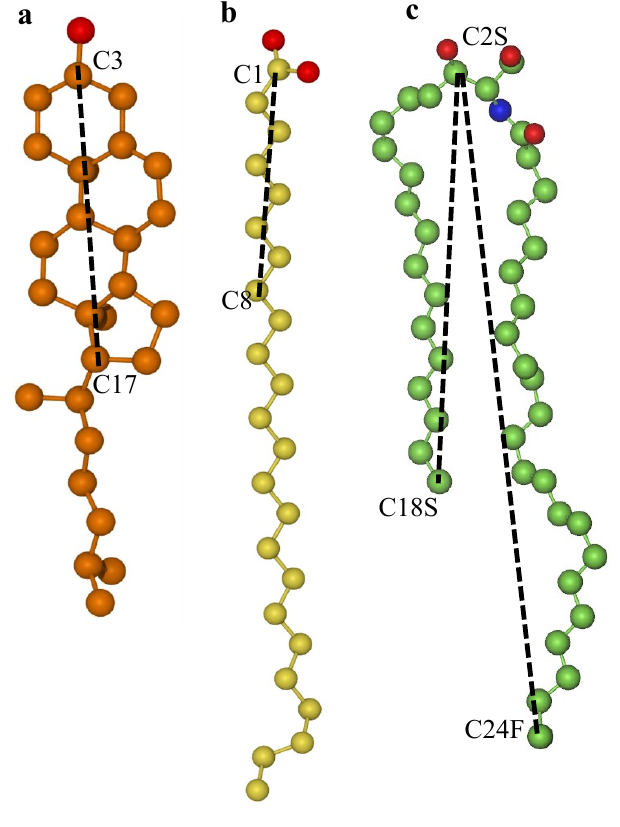}
 \caption{Unit vectors describing molecular orientation are placed along the dashed lines for CHL (a), FFA (b), and CER (c). Only heavy atoms are shown in the structural representations.}
 \label{fig:unit_vectors}
\end{figure}

As foreshadowed by Figure~\ref{fig:flipping_example}, lipids reorient during the flipping process.
To follow this reorientation, we measured the tilt angles of the lipids, defined in terms of the angle between suitably chosen molecular unit vectors placed along the longitudinal axis of the lipids (Figure~\ref{fig:unit_vectors}) and the membrane normal.

\begin{figure}[htb]
 \centering
 \includegraphics[width=\columnwidth]{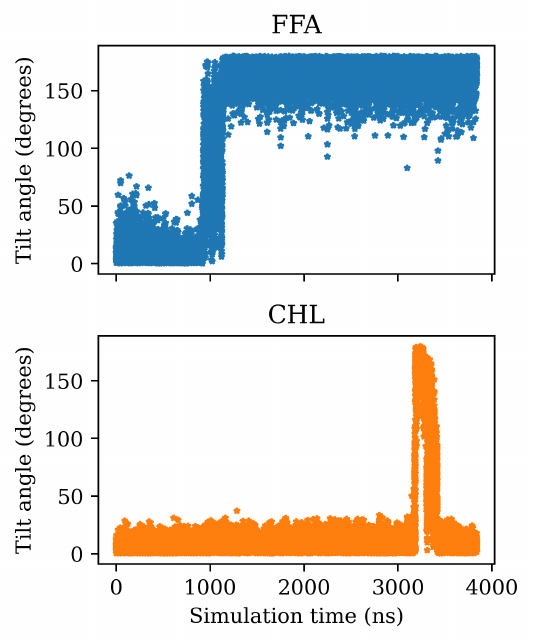}
 \caption{Exemplary tilt angle time series for FFA and CHL in one of the WTM simulations of the SC membrane.}
 \label{fig:tilt_angle_time_series}
\end{figure}

Time series plots of tilt angles show distinct jumps whenever a lipid flips (Figure~\ref{fig:tilt_angle_time_series}).
Through inspection of all tilt angle time series, we determined that only cholesterol and free fatty acid molecules undergo flipping. 
The lack of flipping for ceramides is consistent with the observations from the works of Wang et~al.\cite{wang_models_2018} and Jiang et~al.\cite{jiang_models_2024} for an SPP bilayer. 

Despite \si{\micro\second} simulation times, flipping events were found to be rare and stochastic, and not enough data could be gathered to determine rates or make significant comparisons between systems and methods.
However, we note that the lifetimes of the flipping-induced asymmetries in the system can reach several \si{\micro\second}, as evidenced by the example shown in Figure~\ref{fig:tilt_angle_time_series} (top panel).
We also note that flipping events were observed in the unbiased simulations of the SC membrane, indicating that this is not an artifact of the bias introduced by the enhanced sampling methods but an inherent characteristic of the system under this force field.
We therefore have strong reason to believe that such events play a major role in the slow convergence of the FE profiles described earlier. 

\subsubsection{Contact analysis reveals a significant degree of lipid flipping}

To further characterize the flipping of the lipids, we calculated the number of lipids coordinating to a solute (number of contacts).
We stratified this property by the position of the solute along the bilayer normal (i.e., by the collective variable).
Contacts are defined to exist if the hydroxyl oxygen atoms of the lipids are within a cutoff distance of \qty{4}{\angstrom} of tagged atoms on the solutes (water: central oxygen, acetone, 6-MHO: carbonyl oxygen).
Further details on the computation of the number of contacts are given in the SI.

\begin{figure}[tb]
 \centering
 \includegraphics[width=\columnwidth]{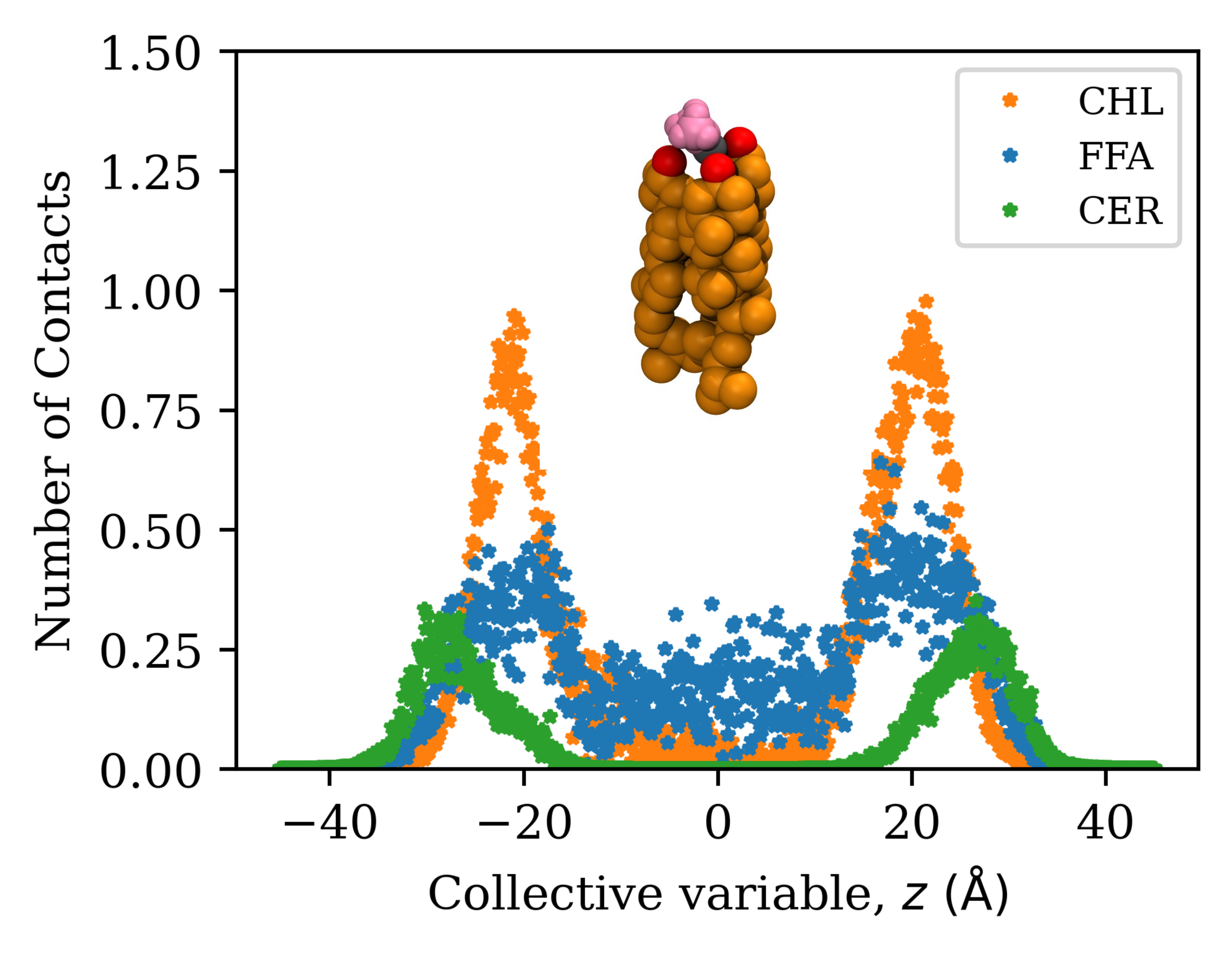}
 \caption{Coordination number between SC lipids and acetone, stratified by the position along the membrane normal. A similar trend was observed for the other solutes (Figure~S9).}
 \label{fig:contacts}
\end{figure}

Maximum coordination between lipids and solutes is expected when the solutes are in the membrane headgroup region around $\pm \qty{28}{\angstrom}$ (compare density profile, Figure~S2).
The coordination number is expected to be zero if the solute is in the center of the membrane and if the membrane maintains its bilayer structure throughout the simulation.
In fact, this is what we observe for the ceramides (Figure~\ref{fig:contacts}).
For cholesterol and free fatty acid molecules, on the other hand, we observe a significant degree of coordination throughout the entire membrane, which indicates that lipids have left their preferred position and orientation in the membrane.
These insights from the coordination number analysis qualitatively align with our observations from the tilt angle analysis.

\subsubsection{Mechanisms of lipid flipping are diverse}

\begin{figure*}[htb]
 \centering
 \includegraphics[width=\textwidth]{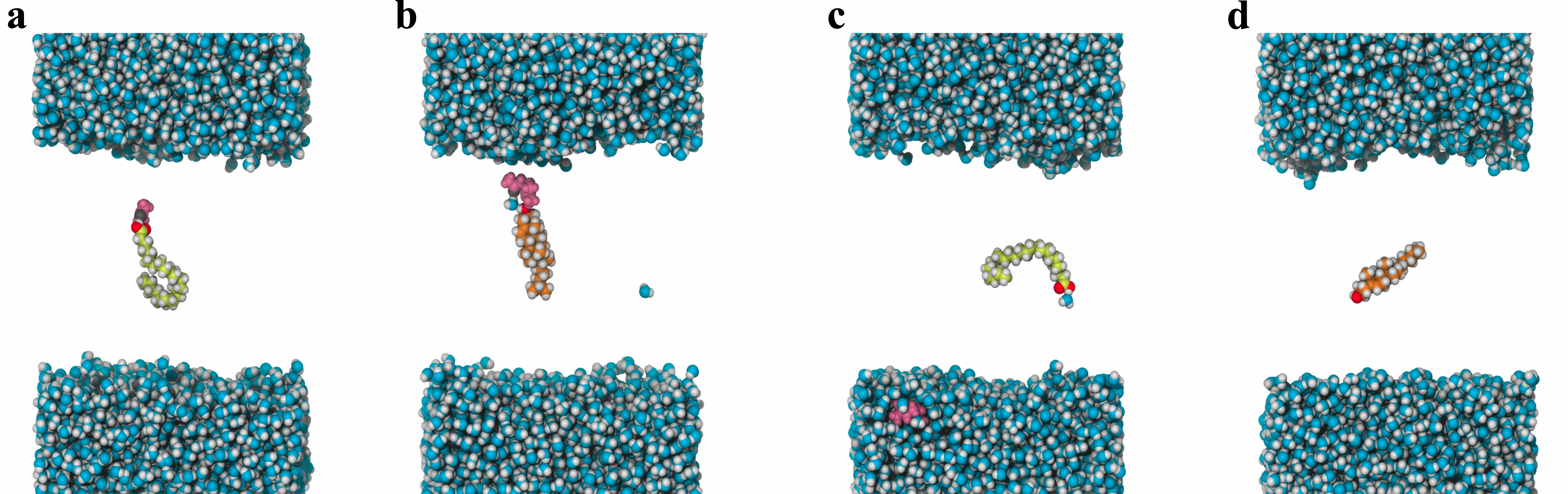}
 \caption{Schematic representation of an acetone-driven flipping of a free fatty acid lipid (a), the flipping of a cholesterol lipid initiated by 6-MHO and with solvent participation (b), the flipping of a free fatty acid lipid driven exclusively by solvent molecules (c), and a cholesterol molecule flipping without the assistance of solvent or solute (d).}
 \label{fig:flipping_mechanisms}
\end{figure*}

By visual inspection of the trajectories, we noticed that the flipping mechanism can be classified into four categories: (a) solute-assisted, (b) solute- and solvent-assisted, (c) solvent-assisted, and (d) unassisted.
In the solute-assisted case, the solute initiates the detachment of the lipid from its initial position (Figure~\ref{fig:flipping_mechanisms}a).
In the solute- and solvent-assisted mechanism, one or more water molecules also participate in the removal of the lipid from its preferred position (Figure~\ref{fig:flipping_mechanisms}b).
In the case of solvent assistance, the conformational change is driven entirely by solvent molecules, without participation of the solute (Figure~\ref{fig:flipping_mechanisms}c). Finally, we also observed that the lipid can move away from the bulk solvent to the middle of the bilayer without the assistance of the solute or solvent molecules (Figure~\ref{fig:flipping_mechanisms}d).
In Table~S1, we provide a list of all flipping events and their classification into different mechanisms.

We note that these mechanisms should be understood as plausible pathways for flipping; the precise nature can only be concluded from running long unbiased simulations, most likely on millisecond time scales or by applying enhanced sampling techniques to the lipids.

\subsection{Permeabilities are sensitive to free energy differences}

The permeability $P$ of a membrane is a macroscopic transport coefficient that connects the particle flux through a membrane to the concentration gradient that causes this flux.
Within the inhomogeneous solubility diffusion (ISD) model,\cite{diamond_interpretation_1974} this quantity can be connected with two microscopically accessible properties, the FE profile~$\Delta{}F(z)$ and the position-dependent diffusivity $D(z)$, and
\begin{equation}
\dfrac{1}{P} =  \int_{-L/2}^{+L/2} \dfrac{e^{\beta \Delta{}F(z)}}{D(z)} \,dz,
\end{equation}
where $L$ is the extent of the membrane. 
Due to exponential dependence, $P$ should be much more sensitive to changes in free energy than to changes in diffusivity, and even minor errors in~$\Delta{}F$ can propagate to large errors in~$P$.
Thus, we investigated how the errors of the FE profiles and insufficient sampling affect the predicted membrane permeabilities.
To focus exclusively on the thermodynamic aspects of permeation and to estimate the effect of methodological errors due to the choice of the enhanced sampling technique, we assume a constant, uniform diffusivity profile, $D(z) = D$, and report reduced permeabilities, 
\begin{equation}
\dfrac{P}{D} = \left[ \int_{-L/2}^{+L/2} {e^{\beta \Delta{}F(z)}} \,dz \right]^{-1}.
\label{eqP2}
\end{equation}

\begin{figure}[tb]
 \centering
 \includegraphics[width=\columnwidth]{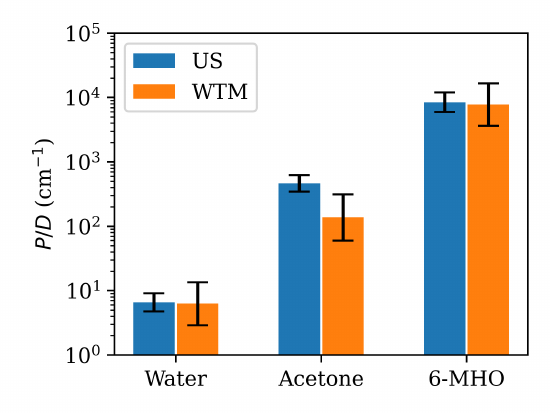}
 \caption{Reduced permeability $P/D$ for all investigated solutes in the SC membrane. Error bars represent \SI{95}{\percent} confidence intervals. Force-symmetrized FE profiles (Figure~\ref{fig:fe_profiles}) were used in the computation.}
 \label{fig:permeability}
\end{figure}

Figure~\ref{fig:permeability} shows that the permeants chosen span a range of magnitudes on the reduced permeability scale $P/D$.
The associated errors depend on the sampling method employed, with US being more accurate than WTM for a given amount of simulation data (Table~\ref{tab:simulation_times}).

Although we do observe nearly quantitative agreement between US and WTM for water and 6-MHO, the permeabilities for acetone differ quite substantially and are at the limit of what can be considered to be equal within error bars.
This was already foreshadowed by the differences in barrier heights (Table~\ref{tab:barrier_heights}).
We note that the WTM error bars for acetone are the largest observed in this study, spanning about an order of magnitude on the reduced-permeability scale.

\begin{figure*}[tb]
 \centering
 \includegraphics[width=\textwidth]{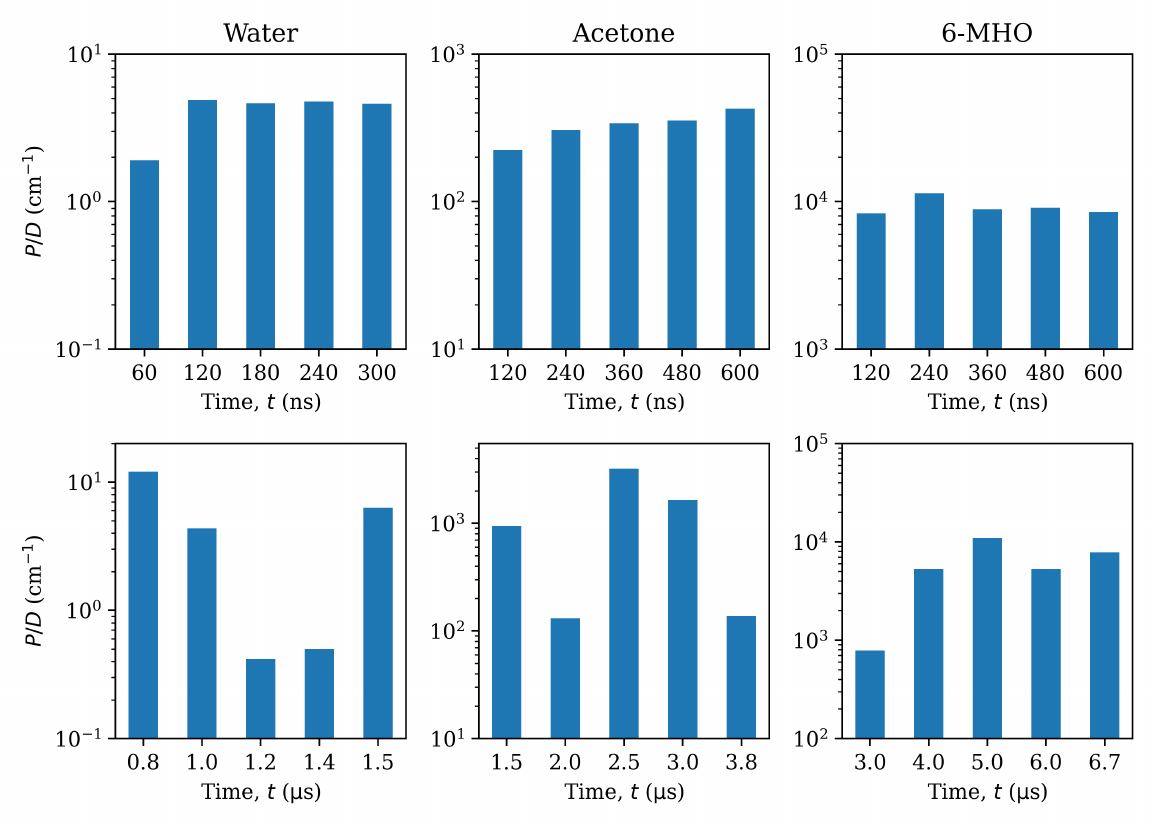}
 \caption{Reduced permeability $P/D$ as a function of simulation time. Top row: US, bottom row: WTM. Force-symmetrized FE profiles were used in the computation. US simulation times are per window. Total simulation times can be obtained by multiplying with the number of windows ($N_\text{win} = 91$). They are also reported in Table~\ref{tab:simulation_times}.}
 \label{fig:convergence_pd}
\end{figure*}

Further insight can be gained by studying the evolution of the reduced permeabilities with simulation time.
Although we do see convergence for US (Figure~\ref{fig:convergence_pd}, top row), where we could presumably have simulated less, this is not the case for WTM (Figure~\ref{fig:convergence_pd}, bottom row).

\section{Conclusions}

In the present study, we carried out MD simulations of three solutes (water, acetone, and 6-MHO) permeating through a stratum corneum model membrane. 
We found that the convergence of the free energy profiles describing permeation toward symmetry is extremely slow, requiring \si{\micro\second} simulation times.
Using two conceptionally different enhanced sampling methods (US and WTM) and by comparing the SC membrane to a conventional fluid-like membrane composed of POPC, we confirmed that the sampling bottleneck is a feature of the SC model and not an artifact of an improperly employed methodology.

Similar sampling challenges have been faced by other authors, for example, Maibaum and Pokhrel, who studied permeation through membranes composed of  1,2-dioleoyl-sn-glycero-3-phosphocholine (DOPC), showing that proper sampling of membrane permeation is challenging, particularly in the case of polar or charged permeants, which can lead to slow relaxation times of electrostatic interactions and to erroneous dependencies of the FE profiles on initial conditions.\cite{pokhrel_free_2018}
In this study, we identified a high degree of order in SC vs POPC and the stochastic flipping of lipids, which induces long-lived structural asymmetries, as likely sources for the slow convergence of free energy profiles.

Dynamical bottlenecks in collective-variable-based free energy calculations can often be overcome by finding more suitable collective variables or by increasing the CV space in order to bias orthogonal degrees of freedom.
Although there is a certain flexibility in the choice of collective variables for studying permeation (by including, for example, variables describing molecular orientation), it is not straightforward to assume that this accelerates the convergence of the free energy profile observed in this study.
Here, convergence is likely hindered by slow reorganization processes of the membrane itself, biasing of which is nontrivial and not possible with readily available implementations of collective variables.

In the final section of the study, we evaluated reduced permeabilities that allow us to focus on the thermodynamic aspects of permeation.
The chosen solutes were shown to span multiple orders of magnitude on the reduced permeability scale.
Although we did observe nearly quantitative agreement between both enhanced sampling methods for water and 6-MHO, differences were noticeable for acetone.
Our results suggest that our initial simulation target of reaching near-symmetry in the FE profiles was likely not sufficient for acetone. The convergence of the force-symmetrized profiles, which we observed for US but not for WTM (Figures~\ref{fig:us_convergence_fe}, \ref{fig:wtm_convergence_fe}), should also be sought.
We presume that this lack of convergence is the ultimate reason for the large error bars for the permeabilities observed for WTM, which reach almost an order of magnitude for acetone.

We would like to emphasize once again that the performance of enhanced sampling techniques is critically dependent on the precise choice of the parameters required by these methods. Examples include the number, placement, and width of the bias potentials in umbrella sampling (US), as well as the hill width, hill height, bias factor, and deposition rate in well-tempered metadynamics (WTM).
In this study, we did not employ a simulation protocol that allows for an unbiased comparison of the efficiency between these two sampling methods. Instead, we benchmarked our results against another system (POPC). Nonetheless, with the setup used in this study, we found that US performed better than WTM in terms of the convergence of our target quantity: permeability. However, this came at the cost of a \numrange{8}{15}-fold increase in total simulation time.
If computational cost is not a concern, US might be the preferred method, as it is inherently parallelizable.

Regardless of the enhanced sampling method chosen, we strongly recommend carefully checking for convergence of free energy profiles in future permeation studies, even if this requires significantly increased simulation times. Sampling only half of the system or eliminating asymmetries through premature symmetrization could lead to severe errors.

We conclude this study by noting that even the largest error interval for the permeabilities observed in this study (acetone, WTM) is likely good enough in the context of indoor air kinetic modeling, where transport coefficients with an exponential dependence on energy are usually considered to be acceptable if they are accurate to within an order of magnitude.
The overall methodology thus remains a valuable tool for constraining parameters needed by kinetic models.

\begin{acknowledgement}

The authors thank Chinmey Das and Peter~D.~Olmsted for helpful discussions about setting up SC models.
DJT thanks the Alfred P. Sloan Foundation Chemistry of the Indoor Environment (CIE) Program (G-2020-13912).
MVD thanks the Deutsche Forschungsgemeinschaft (DFG, German Research Foundation, No. 409294855) for financial support during the early stages of the project.

\end{acknowledgement}

\begin{suppinfo}

The supporting information includes a more detailed explanation of the FE profile error analysis; additional methodological details on the contact number analysis; the density profile of the SC system; additional FE profiles; a discussion of the order parameter profiles; contact number profiles not shown but referenced in the main text;  and an overview of all observed lipid flipping events, their duration and their classification.

\end{suppinfo}

\clearpage
\bibliography{manuscript}

\end{document}

\maketitle

\section{Supplementary Methodological Details}

\subsection{Error analysis}

Standard errors for the FE profiles obtained from US were estimated as described by Hummer et al.,\cite{zhu_convergence_2012}
\begin{equation}
    \mathsf{var}[F(z_n)] \approx \left(\frac{K}{2} \Delta z\right)^2 \cdot \sum_{i=0}^{n} \mathsf{var}(\bar{z}_i),
    \label{eq:us_error}
\end{equation}
where $K$ and $\Delta{}z$ have already been defined in the main text and correspond to the harmonic force constant and the spacing between two adjacent umbrella potentials, respectively.
Furthermore, $\mathsf{var}(\bar{z}_n)$ denotes the squared standard error of the mean of the collective variable $z$ evaluated for the $n$th window.
These were evaluated by applying the automated blocking method described by Jonsson.\cite{Jonsson.Jonsson.2018.StandardErrorEstimation}

\begin{figure}[htb]
 \centering
 \includegraphics[width=\textwidth]{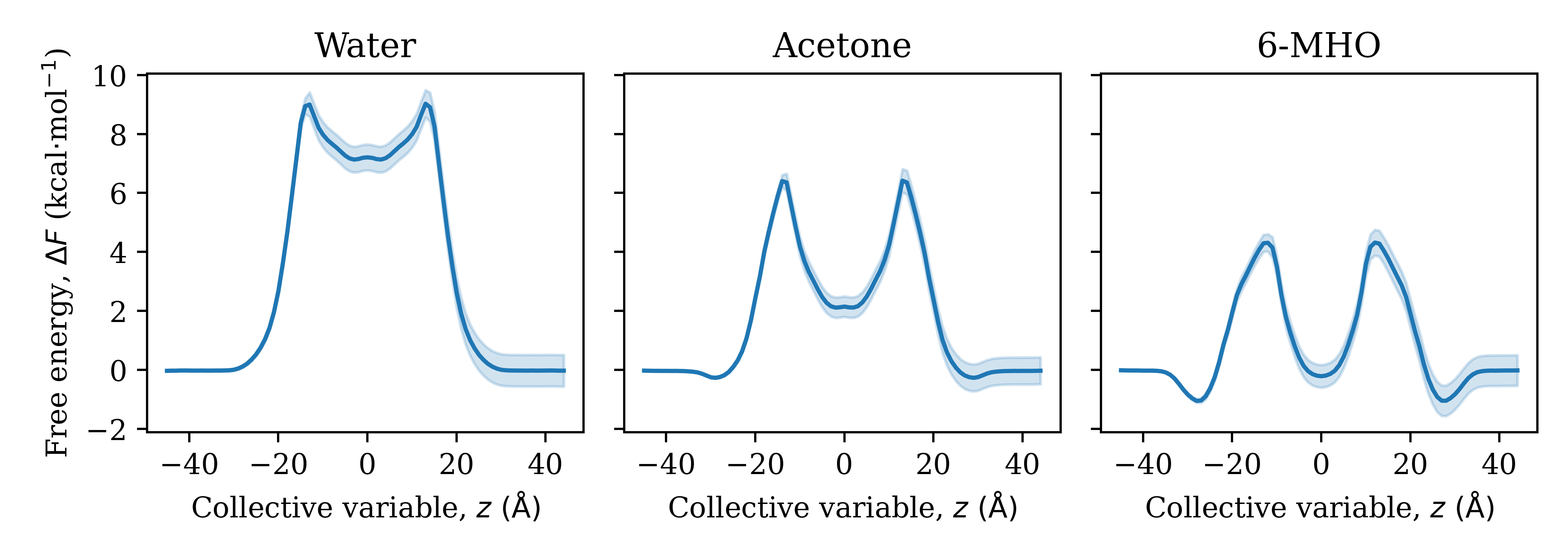}
 \caption{Error propagation through multiple windows in US simulations.}
 \label{fig:cumulative_error}
\end{figure}

On the left-hand side of the equation, $\mathsf{var}[F(z_n)]$ denotes the squared standard error of the free energy evaluated at $z_n$.
Note that errors accumulate over the windows, that is, they are asymmetric by construction (Figure~\ref{fig:cumulative_error}).
This is reasonable since the first window, $n=0$, is chosen as the reference point for the FE profile and cannot have an error by definition.
To obtain a symmetrized version of the error bars, we carried out the same procedure, but treated both halves of the membrane as independent samples.
To be precise, we evaluated Eq.~\ref{eq:us_error}, both for $n=0 \dots n_\mathsf{max}/2$ and for $n=n_\mathsf{max} \dots n_\mathsf{max}/2$, where $n_\mathsf{max}=90$ in our simulations.
We then averaged over pairs of symmetry-equivalent FE values, using the laws for error propagation of uncorrelated random variables,
\begin{align}
    F^\mathsf{sym}(z_n) &= \frac{1}{2}\left[ F\left(z_n\right) + F\left(z_{n_{\mathsf{max}}-n}\right) \right]\,, \label{eq:fe_average}\\
        \mathsf{var}[F^\mathsf{sym}(z_n)] &= \frac{1}{4}\left\lbrace \mathsf{var}[F(z_n)] + \mathsf{var}[F(z_{n_{\mathsf{max}}-n})] \right\rbrace\,.
\end{align}
To obtain the final FE profiles that span the entire CV space, we mirrored the symmetrized free energies and errors with respect to the center of symmetry ($z=0$).

We adopted a similar strategy for the WTM simulations.
Here, snapshots of the FE profiles were saved every \SI{2}{\nano\second}.
Each of these FE profiles was bifurcated with respect to the center of symmetry and referenced so that the FE in the center of the aqueous region is $0$.
Finally, Jonsson's automated blocking method was applied to each half FE profile and averages were computed as described above.

\subsection{Number of Contacts}

The number of contacts $N$ between a solute $(s)$ and the lipids $(l)$ was calculated using 
\begin{equation}
    N =  \sum_{l} \frac{1-\left(\left| \mathbf{r}_s - \mathbf{r}_{l}\middle|\right/d_c\right)^{n}}{1-\left(\left| \mathbf{r}_s - \mathbf{r}_{l}\middle|\right/d_c\right)^{m}}\,.
    \label{eq:contacts}
\end{equation}
Here, $\mathbf{r}_s$ refers to the position of the oxygen atoms of the solutes (for water: hydroxyl, for acetone, 6-MHO: carbonyl), $\mathbf{r}_{l}$ denotes the position of the hydroxyl oxygen atoms of the lipids, $d_c = \qty{4}{\angstrom}$ is the cutoff distance, and $n = 6$ and $m = 12$ are common parameter choices that ensure that the function decays to zero quickly for separations larger than $d_c$.
The sum runs over all lipids.
For ceramides, which have two hydroxyl groups, we present averages.

\section{Supplementary Results}

\subsection{Density profile}

\begin{figure}[H]
 \centering
 \includegraphics[width=3.3in]{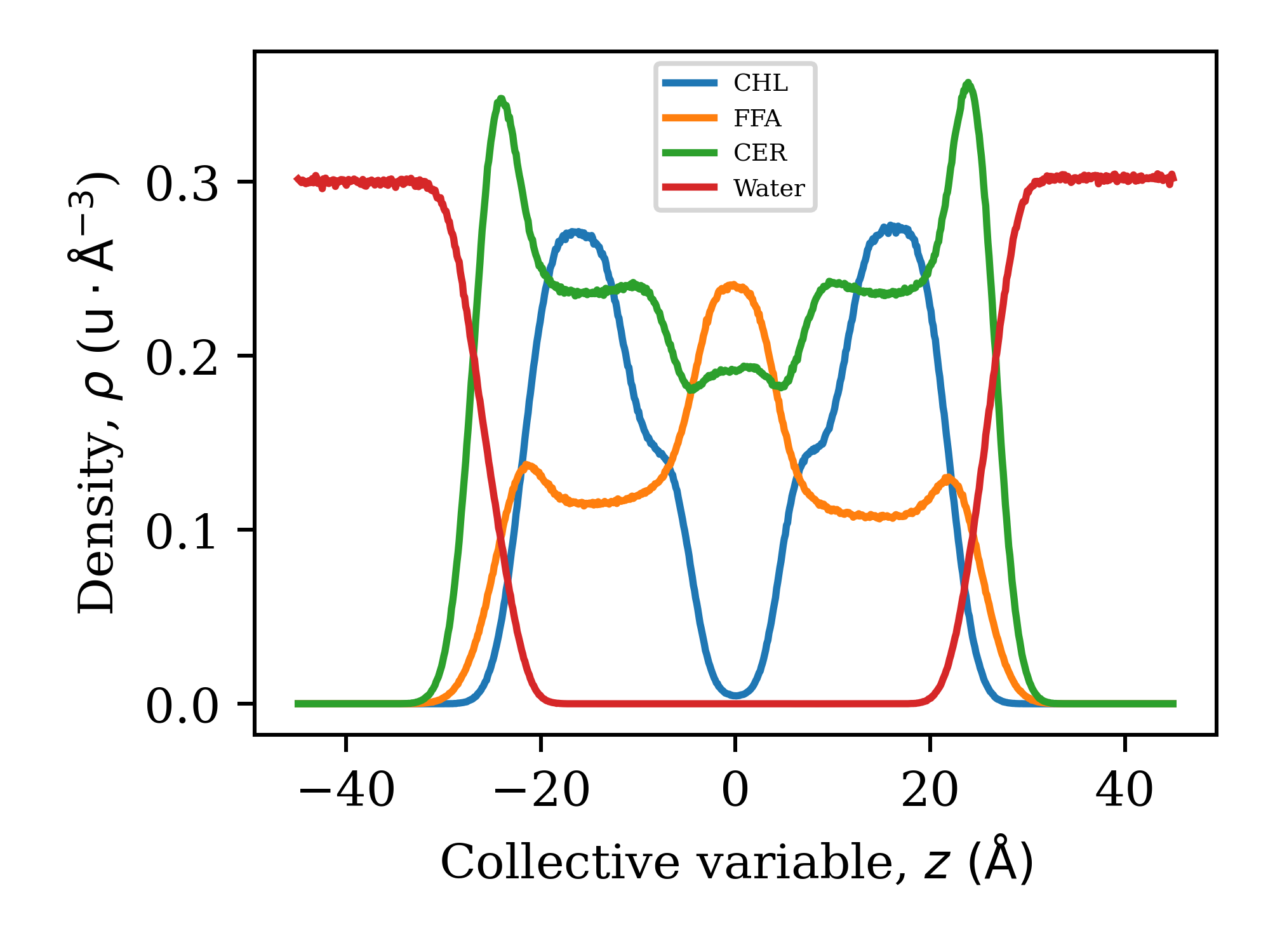}
 \caption{Density profiles of the FFA, CHL, CER and water in the SC system.}
 \label{fig:figureS2_density_profile}
\end{figure}

We computed the mass density profile $\rho$ in bins of size $\Delta{}z=\SI{0.1}{\angstrom}$ along the collective variable for the SC bilayer (Figure~\ref{fig:figureS2_density_profile}).
The profile is perfectly symmetric, despite the lipid flipping described in the main text. It permits an easy identification of the aqueous phase, the lipid headgroup region, and the tails.

\subsection{Free energy profiles}

\begin{figure}[htbp]
    \centering
    \includegraphics[width=\textwidth]{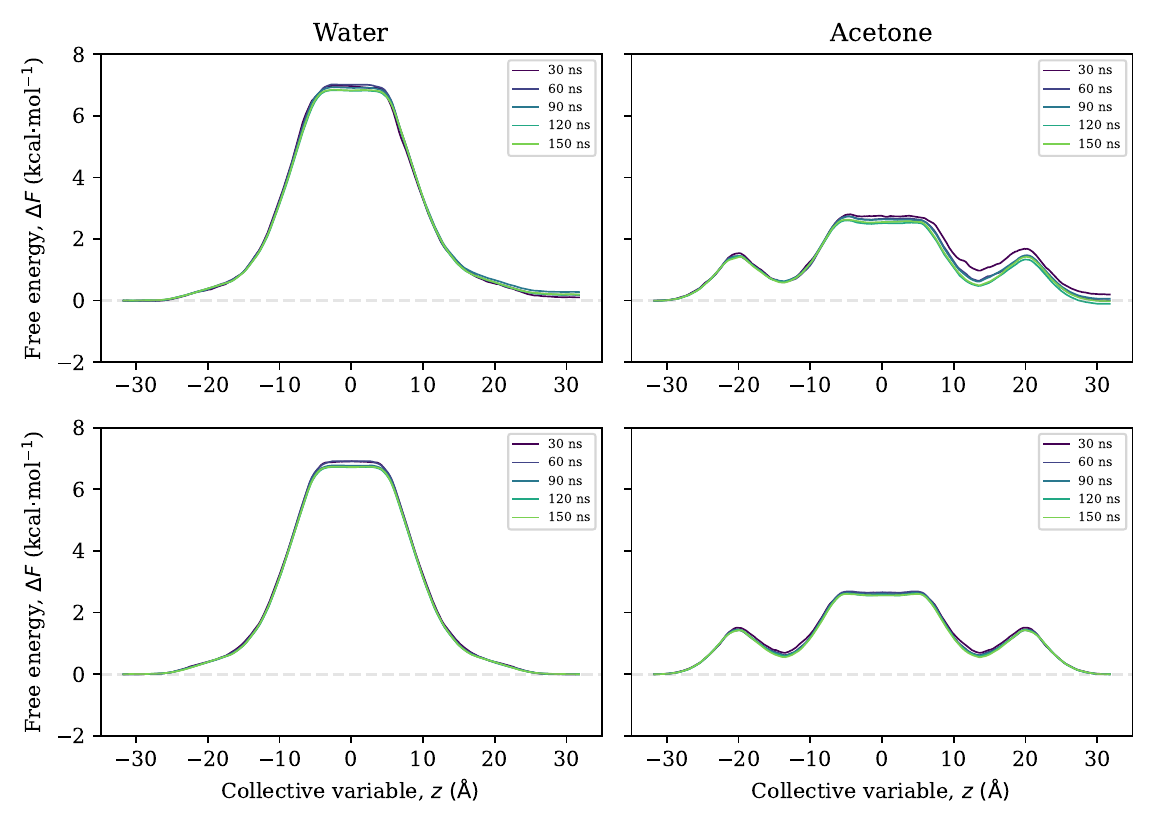}
    \caption{Top row: Convergence of POPC FE profiles toward symmetry for US. Bottom row: Convergence of the corresponding force-symmetrized FE profiles toward their final values. All simulation times are per window. The total simulation times, which we report in Table~1, are in the \si{\micro\second} region.}
    \label{fig:us_convergence_fe_popc}
\end{figure}

\begin{figure}[htbp]
    \centering
    \includegraphics[width=\textwidth]{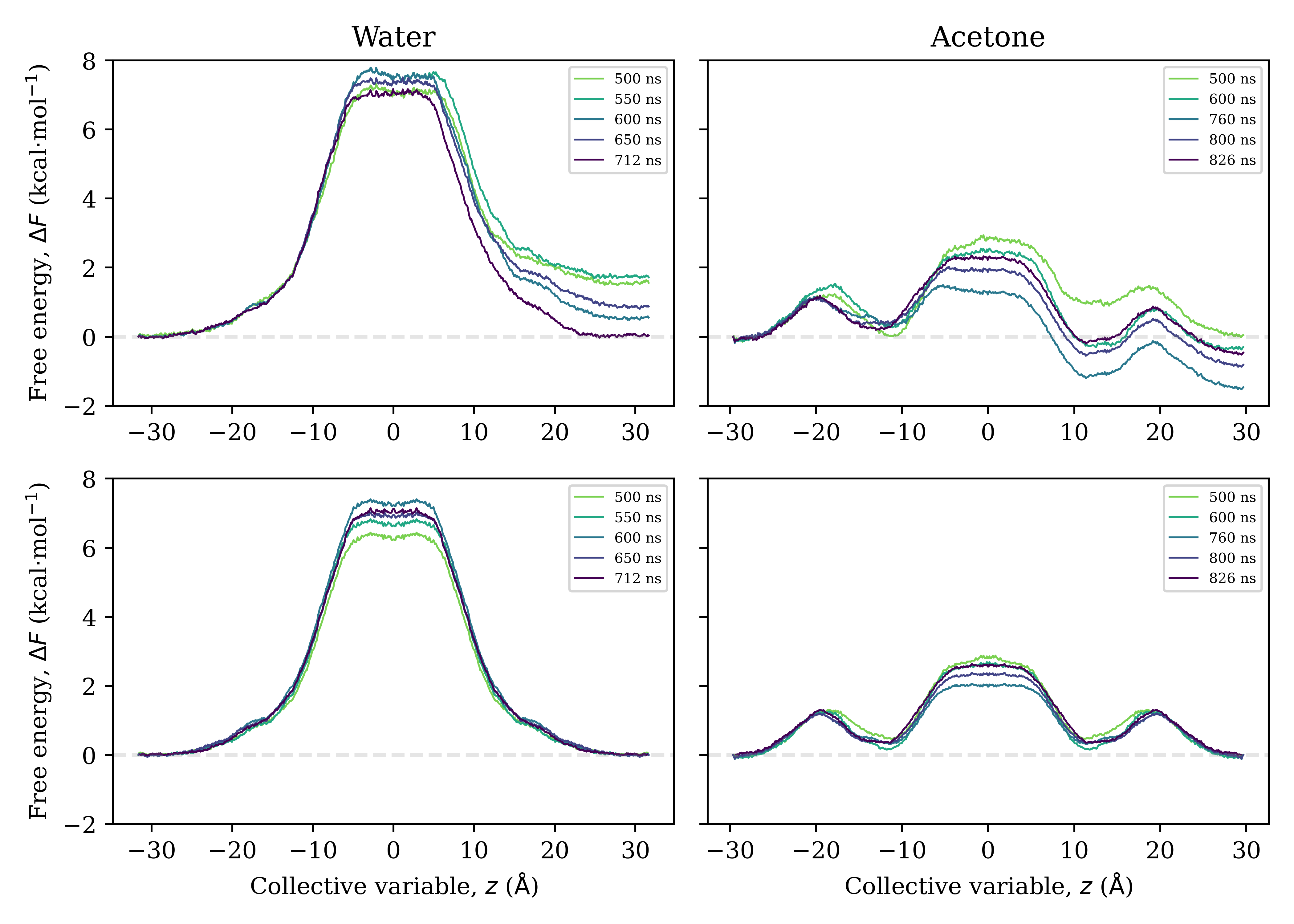}
    \caption{Top row: Convergence of POPC FE profiles toward symmetry for WTM. Bottom row: Convergence of the corresponding force-symmetrized FE profiles toward their final values.}
    \label{fig:wtm_convergence_fe_popc}
\end{figure}

\begin{figure}[htbp]
    \centering
    \includegraphics[width=3.3in]{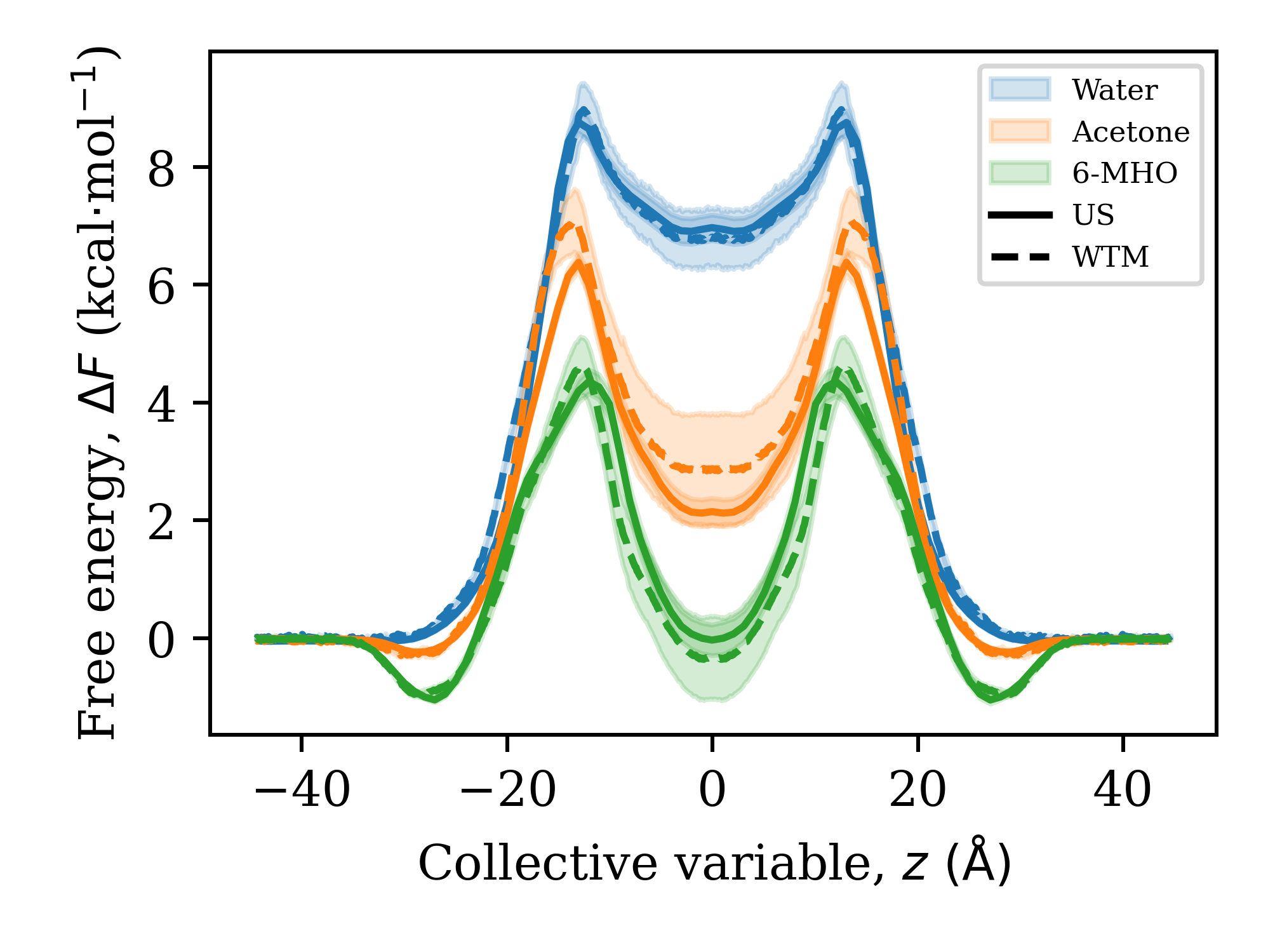}
    \caption{Final, force-symmetrized FE profiles for all investigated solutes in the SC model. Unsmoothened version of Figure~5 shown in the main text.}
    \label{fig:sc_pmfs_raw}
\end{figure}

\begin{figure}[htbp]
   \centering
    \includegraphics[width=\textwidth]{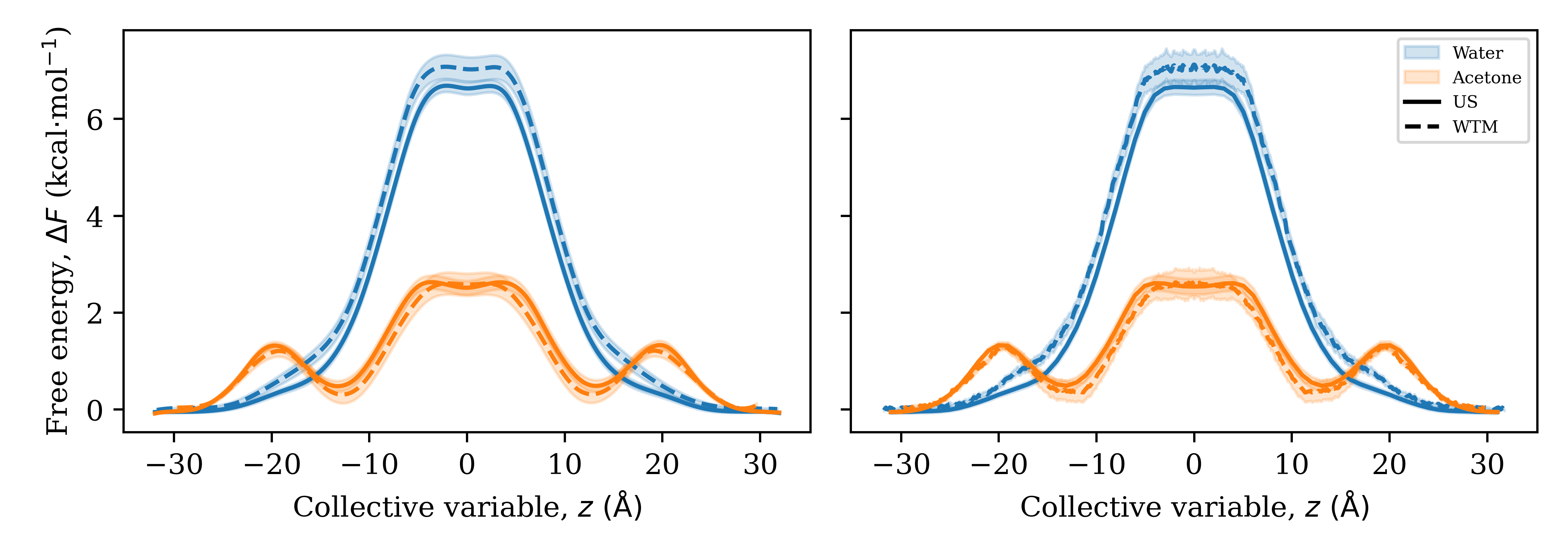}
    \caption{Final, force-symmetrized FE profiles for all investigated solutes in the POPC model with smoothened version in the left panel and the unsmoothened version in the right panel}
    \label{fig:popc_pmfs}
\end{figure}

Figures~\ref{fig:us_convergence_fe_popc} and Figure~\ref{fig:wtm_convergence_fe_popc} are the equivalents of Figures~3 and~4 for POPC, showing the convergence toward symmetry of the FE profiles.
In Figure~\ref{fig:sc_pmfs_raw}, we give the final force-symmetrized PMFs for the SC system in its raw version, without using smoothing.
In Figure~\ref{fig:popc_pmfs} we give the final force-symmetrized PMFs for the POPC system.

For acetone, we find qualitative differences to similar results reported in the literature, which are likely due to differences in the force fields used to model the interactions in the system (Gromos / OPLS / SPC vs Charmm / CGenFF / TIP3P).\cite{posokhov_effect_2013}.
However, we note that our results are qualitatively in agreement with the octanol--water partition coefficient reported by Cumming et al.\cite{cumming_octanolwater_2017} $(\log{K_\text{OW}} = -0.22)$, indicating that it is thermodynamically unfavorable for acetone to partition from water to a lipophilic region.

\clearpage
\subsection{Order parameters}

To understand the dip in the order parameters observed in the oleoyl chain of POPC and the sphingosine chain of the ceramide lipids, we present here distributions of the cosine of the angle formed by methylene bonds and the bilayer normal. Averaging over the second Legendre polynomial of this cosine gives the order parameter,
\begin{equation}
    S_{\ce{CH}} = \frac{\left<3 \cos^2 \theta - 1\right>}{2}\,.
\end{equation}

\begin{figure}[htbp]
 \centering
 \includegraphics[width=\textwidth]{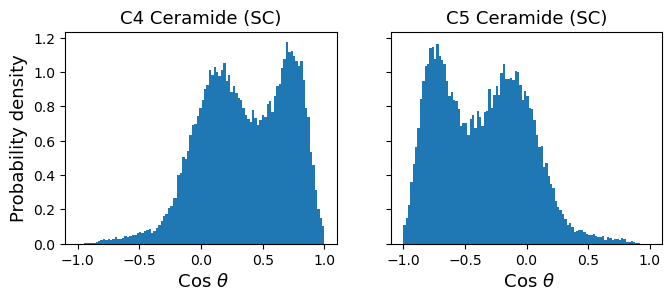}
 \caption{Distribution of the cosine of the angle formed by the sphingosine methylene bonds (located at C4 and C5) and the bilayer normal.}
 \label{fig:OP_SC}
\end{figure}

\begin{figure}[htbp]
 \centering
 \includegraphics[width=\textwidth]{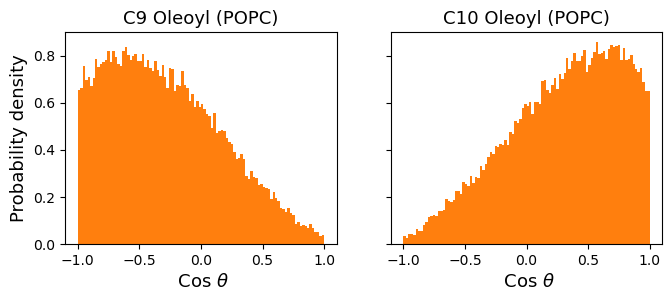}
 \caption{Distribution of the cosine of the angle formed by the oleoyl methylene bonds (located at C9 and C10) and the bilayer normal.}
 \label{fig:OP_POPC}
\end{figure}

To construct the plots shown in Figures~\ref{fig:OP_SC} and \ref{fig:OP_POPC}, the same trajectory was analyzed as used to calculate the order parameters (Figure~6 in the main text). For better clarity, only the lipids in the upper leaflet of the membrane were considered. The data points were grouped into 100 bins of equal width, and the final distribution was normalized to provide the probability density.

While the methylene bonds adopt a range of orientations in both systems, the distributions are not uniform but are centered at $\cos(\theta) \approx \pm 0.57$, which corresponds to the "magic" angle, for which $S_{\ce{CH}} = 0$. These findings explain why the order parameter nearly vanishes for double-bonded carbon atoms in POPC ($|S_{\ce{CH}}| \approx 0.05$) and why it has a relatively low value in sphingosine ($|S_{\ce{CH}}| \approx 0.13$).

We note that similar findings have been reported by other authors.\cite{Seelig.Waespe-Sarcevic.1978.MolecularOrderCisa,Niemela.Vattulainen.2004.StructureDynamicsSphingomyelin,Merz.Merz.1997.MolecularDynamicsSimulations}

\subsection{Number of contacts}

\begin{figure}[H]
 \centering
 \includegraphics[width=\textwidth]{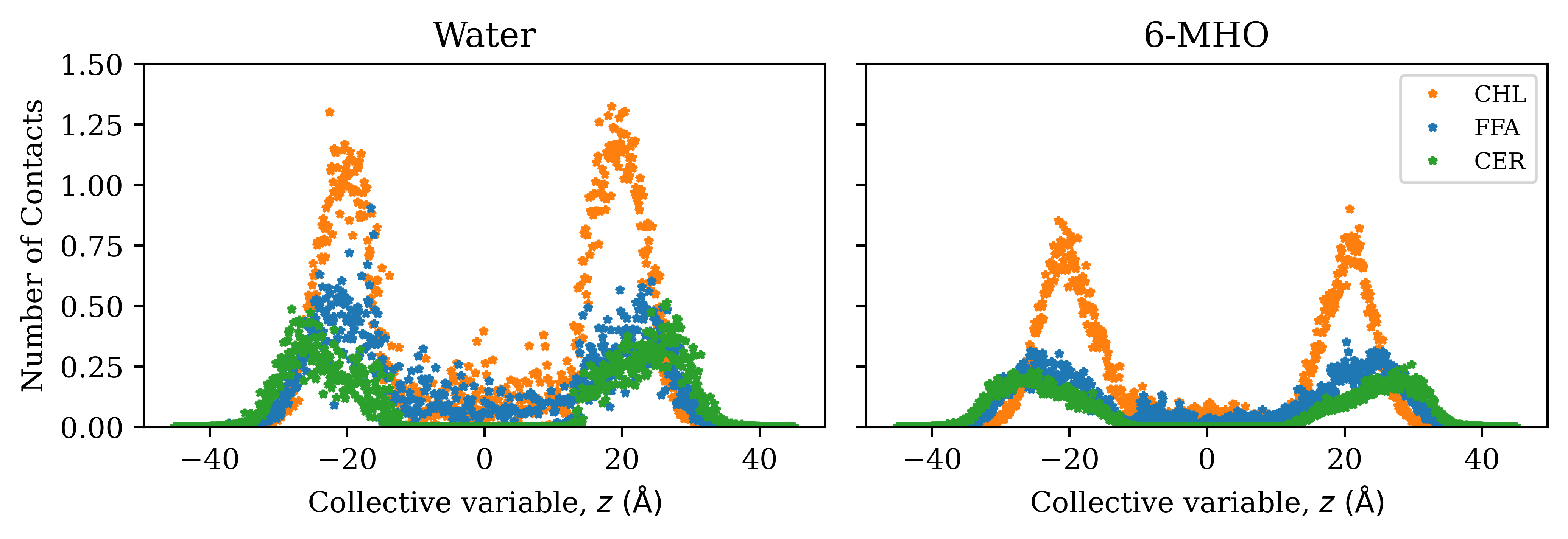}
 \caption{Average coordination number between SC lipids and water (left panel) as well as 6-MHO (right panel), stratified by the position along the membrane normal.}
 \label{fig:wtm_water_6mho_contacts}
\end{figure}

\subsection{Flipping events and mechanisms}

\begin{table}[H]
\centering
  \caption{Summary of all flipping events observed in the WTM and unbiased simulations. The mechanisms a-d have are described as follows; (a) solute assisted, (b) solute and solvent assisted (c) solvent assisted and (d) unassisted.}
  \begin{tabular*}{\textwidth}{@{\extracolsep{\fill}}llS[table-format=4.0]S[table-format=4.0]c}
    \toprule
    System & Residue & {Start (\si{ns})} & {Stop (\si{ns})} & {Mechanism} \\
    \midrule
    Water in SC      & CHL  & 49   & 367  & d \\
    Water in SC      & CHL  & 843  & 1113 & b \\
    Water in SC      & FFA  & 122  & 125  & d \\
    Water in SC      & FFA  & 371  & 393  & b \\
    Acetone in SC    & CHL  & 3177 & 3419 & d \\
    Acetone in SC    & FFA  & 45   & 48   & a \\
    Acetone in SC    & FFA  & 100  & 115  & a \\
    Acetone in SC    & FFA  & 646  & 704  & a \\
    Acetone in SC    & FFA  & 928  & 1132 & a \\
    Acetone in SC    & FFA  & 1717 & 1998 & a \\
    6-MHO in SC      & CHL  & 61   & 119  & a \\
    6-MHO in SC      & CHL  & 197  & 502  & a \\
    6-MHO in SC      & CHL  & 836  & 1035 & b \\
    6-MHO in SC      & CHL  & 1097 & 1172 & a \\
    6-MHO in SC      & CHL  & 5967 & 5978 & a \\
    6-MHO in SC      & FFA  & 419  & 565  & b \\
    6-MHO in SC      & FFA  & 4327 & 4382 & c \\
    6-MHO in SC      & FFA  & 6517 & 6657 & d \\
    Unbiased Neat SC & CHL  &  33  & 199  & d \\
    \bottomrule
    \label{Summary_flip}
  \end{tabular*}
\end{table}

\bibliography{si}